\documentclass[12 pt,twoside,aps,prd,amsmath,amssymb,
tightenlines,showpacs,showkeys,eqsecnum]{revtex4}

\usepackage{graphicx}
\usepackage{mathptmx}
\usepackage{bm}
\newcommand {\ve}{\varepsilon}

\newcommand {\cD}{\cal D}

\newcommand {\bp}{\bar \psi}

\def\myfigure #1#2#3#4
{\begin{figure}[ht]\begin{center}
\includegraphics[width=#2 \textwidth]{#1.eps}
\parbox[t]{#4\textwidth}{\caption{#3}\label{#1}}
\end{center}\end{figure}}

\def \myfigures #1#2#3#4#5#6#7#8
{\begin{figure}[ht]
    \begin{center}
        \includegraphics[width=#2 \textwidth]{#1.eps}
        \hfill
        \includegraphics[width=#6 \textwidth]{#5.eps}
        \parbox[t]{#4\textwidth}{\caption {#3}\label{#1}}
        \hfill
        \parbox[t]{#8\textwidth}{\caption {#7}\label{#5}}
    \end{center}
\end{figure} }

\begin{document}
\baselineskip -24pt
\title{Spinor model of a perfect fluid: examples}
\author{Bijan Saha}
\affiliation{Laboratory of Information Technologies\\
Joint Institute for Nuclear Research, Dubna\\
141980 Dubna, Moscow region, Russia} \email{bijan@jinr.ru}
\homepage{http://wwwinfo.jinr.ru/~bijan/}

\begin{abstract}

Different characteristic of matter influencing the evolution of the
Universe has been simulated by means of a nonlinear spinor field.
Using the nonlinear spinor field corresponding to different types of
perfect fluid or dark energy as a source of gravitational field,
evolution of a Bianchi type-I universe has been illustrated.

\end{abstract}



\maketitle

\bigskip

\section{Introduction}
One of the principal goal of cosmological models is the description
of the different phases of the Universe. In doing so researchers use
different types of sources from perfect fluid to dark energy. In
recent time many authors have studied the evolution of the Universe
where the source is given by a nonlinear spinor field
\cite{henprd,sahaprd,greene,SBprd04,kremer1,ECAA06,sahaprd06,BVI}.
In those papers it was shown that a suitable choice of nonlinearity
(i) accelerates the isotropization process, (ii) gives rise to a
singularity-free Universe and (iii) generates late time
acceleration. In a recent paper \cite{shikin} the authors have
simulated perfect fluid using spinor field with different
nonlinearity. That paper was followed by one \cite{spinpf0}, where
perfect fluid and dark energy were modeled by nonlinear spinor
field. In doing so we used two types of nonlinearity, one occurs as
a result of self-action and the other resulted from the interaction
between the spinor and scalar field. It was shown that the case with
induced nonlinearity is the partial one and can be derived from the
case with self-action. So, in this paper we stick to the case with
self-action. Here we repeat some of the previous results, give the
description of generalized Chaplygin gas and  modified quintessence
\cite{pfdenr} in terms of spinor field and study the evolution of
the Universe filled with nonlinear spinor field within the scope of
a Bianchi type-I cosmological model.

\section{Simulation of perfect fluid with nonlinear spinor field}

First of all let us note that one of the simplest and popular model
of the Universe is a homogeneous and isotropic one filled with a
perfect fluid with the energy density $\ve = T_0^0$ and pressure $p
= - T_1^1 = -T_2^2 = -T_3^3$ obeying the barotropic equation of
state
\begin{equation}
p = W \ve, \label{beos}
\end{equation}
where $W$ is a constant. Depending on the value of $W$ \eqref{beos}
describes perfect fluid from phantom to ekpyrotic matter, namely
\begin{subequations}
\label{zeta}
\begin{eqnarray}
W &=& 0, \qquad \qquad \qquad {\rm (dust)},\\
W &=& 1/3, \quad \qquad \qquad{\rm (radiation)},\\
W &\in& (1/3,\,1), \quad \qquad\,\,{\rm (hard\,\,Universe)},\\
W &=& 1, \quad \qquad \quad \qquad {\rm (stiff \,\,matter)},\\
W &\in& (-1/3,\,-1), \quad \,\,\,\,{\rm (quintessence)},\\
W &=& -1, \quad \qquad \quad \quad{\rm (cosmological\,\, constant)},\\
W &<& -1, \quad \qquad \quad \quad{\rm (phantom\,\, matter)},\\
W &>& 1, \quad \qquad \quad \qquad{\rm (ekpyrotic\,\, matter)}.
\end{eqnarray}
\end{subequations}
In order to describe the matter given by \eqref{zeta} with a spinor
field let us now write the corresponding Lagrangian \cite{sahaprd}:
\begin{equation}
L_{\rm sp} = \frac{i}{2} \biggl[\bp \gamma^{\mu} \nabla_{\mu} \psi-
\nabla_{\mu} \bar \psi \gamma^{\mu} \psi \biggr] - m\bp \psi + F,
\label{lspin}
\end{equation}
where the nonlinear term $F$ describes the self-action of a spinor
field and can be presented as some arbitrary functions of invariant
generated from the real bilinear forms of a spinor field. For
simplicity we consider the case when $F = F(S)$ with $S = \bp \psi$.
We consider the case when the spinor field depends on $t$ only. In
this case for the components of energy-momentum tensor we find
\begin{subequations}
\begin{eqnarray}
T_0^0 &=& mS - F, \label{t00s}\\
T_1^1 = T_2^2 = T_3^3 &=& S \frac{dF}{dS} - F. \label{t11s}
\end{eqnarray}
\end{subequations}

Inserting  $\ve = T_0^0$ and $p = - T_1^1$ into \eqref{beos} we find

\begin{equation}
S \frac{dF}{dS} - (1+W)F + m W S= 0, \label{eos1s}
\end{equation}
with the solution
\begin{equation}
F = \lambda S^{1+W} + mS, \label{sol1}
\end{equation}
with $\lambda$ being an integration constant. Inserting \eqref{sol1}
into \eqref{t00s} we find that
\begin{equation}
T_0^0 = - \lambda S^{1+W}. \label{lambda}
\end{equation}
Since energy density should be non-negative, we conclude that
$\lambda$ is a negative constant, i.e., $\lambda = - \nu$, with
$\nu$ being a positive constant. So finally we can write the
components of the energy momentum tensor
\begin{subequations}
\begin{eqnarray}
T_0^0 &=& \nu S^{1+W}, \label{t00sf}\\
T_1^1 = T_2^2 = T_3^3 &=& - \nu W S^{1+W}. \label{t11sf}
\end{eqnarray}
\end{subequations}
As one sees, the energy density $\ve = T_0^0$ is always positive,
while the pressure $p = - T_1^1 = \nu W S^{1+W}$ is positive for $W
> 0$, i.e., for usual fluid and negative for $W < 0$, i.e. for dark
energy.

In account of it the spinor field Lagrangian now reads
\begin{equation}
L_{\rm sp} = \frac{i}{2} \biggl[\bp \gamma^{\mu} \nabla_{\mu} \psi-
\nabla_{\mu} \bar \psi \gamma^{\mu} \psi \biggr] - \nu S^{1+W},
\label{lspin1}
\end{equation}
Thus a massless spinor field with the Lagrangian \eqref{lspin1}
describes perfect fluid from phantom to ekpyrotic matter. Here the
constant of integration $\nu$ can be viewed as constant of
self-coupling. A detailed analysis of this study was given in
\cite{shikin}.

Let us now generate a Chaplygin gas by means of a spinor field. A
Chaplygin gas is usually described by a equation of state
\begin{equation}
p = -A/\ve^\gamma. \label{chap}
\end{equation}
Then in case of a massless spinor field for $F$ one finds
\begin{equation}
\frac{(-F)^\gamma d(-F)}{(-F)^{1+\gamma} - A} = \frac{dS}{S},
\label{eqq}
\end{equation}
with the solution
\begin{equation}
-F = \bigl(A + \lambda S^{1+\gamma}\bigr)^{1/(1+\gamma)}.
\label{chapsp}
\end{equation}
On account of this for the components of energy momentum tensor we
find
\begin{subequations}
\begin{eqnarray}
T_0^0 &=& \bigl(A + \lambda S^{1+\gamma}\bigr)^{1/(1+\gamma)}, \label{edchapsp}\\
T_1^1 = T_2^2 = T_3^3 &=& A/\bigl(A + \lambda
S^{1+\gamma}\bigr)^{\gamma/(1+\gamma)}. \label{prchapsp}
\end{eqnarray}
\end{subequations}
As was expected, we again get positive energy density and negative
pressure.

Thus the spinor field Lagrangian corresponding to a Chaplygin gas
reads
\begin{equation}
L_{\rm sp} = \frac{i}{2} \biggl[\bp \gamma^{\mu} \nabla_{\mu} \psi-
\nabla_{\mu} \bar \psi \gamma^{\mu} \psi \biggr] - \bigl(A + \lambda
S^{1+\gamma}\bigr)^{1/(1+\gamma)}. \label{lspin2}
\end{equation}
Setting $\gamma = 1$ we find the result obtained in \cite{spinpf0}.

Finally, we simulate modified quintessence with a nonlinear spinor
field. It should be noted that one of the problems that face models
with dark energy is that of eternal acceleration. In order to get
rid of that problem quintessence with a modified equation of state
was proposed which is given by \cite{pfdenr}
\begin{equation}
p = - W (\ve - \ve_{\rm cr}), \quad W \in (0,\,1), \label{mq}
\end{equation}
Here $\ve_{\rm cr}$ some critical energy density.  Setting $\ve_{\rm
cr} = 0$ one obtains ordinary quintessence. It is well known that as
the Universe expands the (dark) energy density decreases. As a
result, being a linear negative function of energy density, the
corresponding pressure begins to increase. In case of an ordinary
quintessence the pressure is always negative, but for a modified
quintessence as soon as $\ve_{\rm q}$ becomes less than the critical
one, the pressure becomes positive.

Inserting $\ve = T_0^0$ and $p = - T_1^1$ into \eqref{mq} we find
\begin{equation}
F = - \eta S^{1-W} + mS + \frac{W}{1-W}\ve_{\rm cr},  \label{Fmq}
\end{equation}
with $\eta$ being a positive constant. On account of this for the
components of energy momentum tensor we find
\begin{subequations}
\begin{eqnarray}
T_0^0 &=& \eta S^{1-W} - \frac{W}{1-W}\ve_{\rm cr}, \label{edmq}\\
T_1^1 = T_2^2 = T_3^3 &=& \eta  W S^{1-W} - \frac{W}{1-W}\ve_{\rm
cr}. \label{prmq}
\end{eqnarray}
\end{subequations}

We see that a nonlinear spinor field with specific type of
nonlinearity can substitute perfect fluid and dark energy, thus give
rise to a variety of evolution scenario of the Universe.

\section{Anisotropic cosmological models with a spinor field}

In the previous two sections we showed that the perfect fluid and
the dark energy can be simulated by a nonlinear spinor field. In the
section II the nonlinearity was the subject to self-action, while in
section III the nonlinearity was induced by a scalar field. It was
also shown the in our context the results of section III is some
special cases those of section II. Taking it into mind we study the
evolution an anisotropic Universe filled with a nonlinear spinor
field given by the Lagrangian \eqref{lspin}, with the nonlinear term
$F$ is given by \eqref{sol1} of \eqref{chapsp}.

We consider the anisotropic Universe given by the Bianchi type-I
(BI) space-time
\begin{eqnarray}
ds^2 = dt^2 - a_1^2 dx^2 -  a_2^2 dy^2 - a_3^2 dz^2, \label{BI}
\end{eqnarray}
with $a_i$ being the functions of $t$ only. From the spinor field
equation, it can be shown that \cite{sahaprd}
\begin{equation}
S = \frac{C_0}{\tau}, \label{S}
\end{equation}
where we define
\begin{eqnarray}
\tau =  \sqrt{-g} = a_1 a_2 a_3. \label{taudef}
\end{eqnarray}
For the components of the spinor field we obtain
\begin{eqnarray}
\psi_{1,2}(t) = \frac{C_{1,2}}{\sqrt{\tau}}\,e^{i\int {\cD}
dt},\quad \psi_{3,4}(t) = \frac{C_{3,4}}{\sqrt{\tau}}\,e^{-i\int
{\cD} dt},
\end{eqnarray}
where ${\cD} = dF/dS$. Solving the Einstein equation for the metric
functions one find \cite{sahaprd}
\begin{eqnarray}
a_i = D_i \tau^{1/3} \exp{\Bigl(X_i \int \frac{dt}{\tau}\Bigr)}
\end{eqnarray}
with the constants $D_i$ and $X_i$ obeying
\begin{equation}
\prod_{i=1}^{3} D_i = 1, \quad     \sum_{i=1}^{3} X_i = 0.
\end{equation}
Thus the components of the spinor field and metric functions are
expressed in terms of $\tau$. From the Einstein equations one finds
the equation for $\tau$ \cite{sahaprd}
\begin{eqnarray}
\frac{\ddot \tau}{\tau}= \frac{3}{2}\kappa
\Bigl(T_{1}^{1}+T_{0}^{0}\Bigr). \label{dtau}
\end{eqnarray}
In case of \eqref{lspin1} on account of \eqref{S} Eq. \eqref{dtau}
takes the form
\begin{equation}
\ddot \tau = (3/2) \kappa \nu C_0^{1+W} (1-W) \tau^{-W}
\end{equation}
with the solution in quadrature
\begin{equation}
\frac{d\tau}{\sqrt{3 \kappa \nu C_0^{1+W} \tau^{1-W} + C_1}} = t +
t_0.
\end{equation}
Here $C_1$ and $t_0$ are the integration constants.

\myfigures{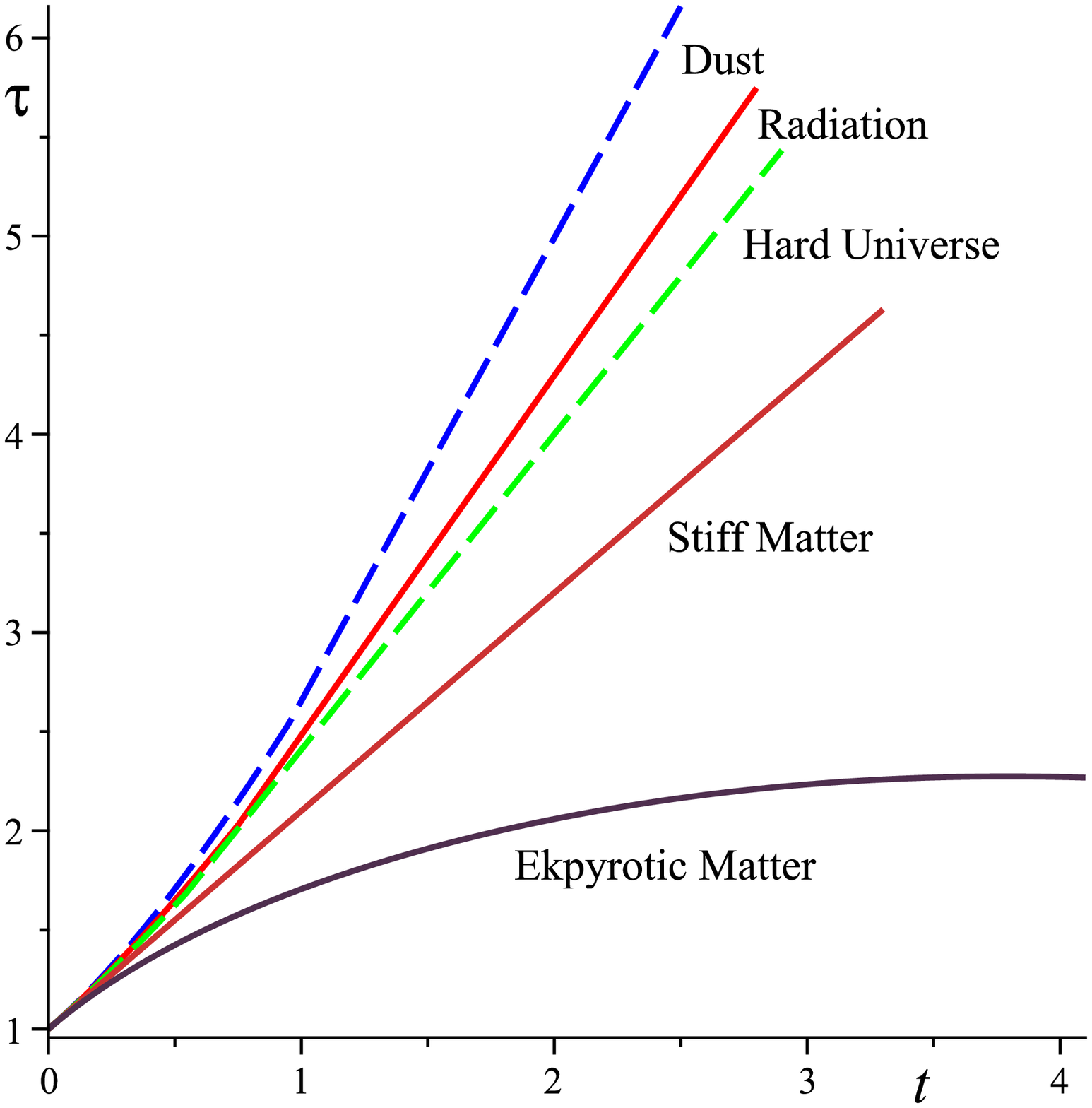}{0.45}{Evolution of the Universe filled with
perfect fluid.} {0.45}{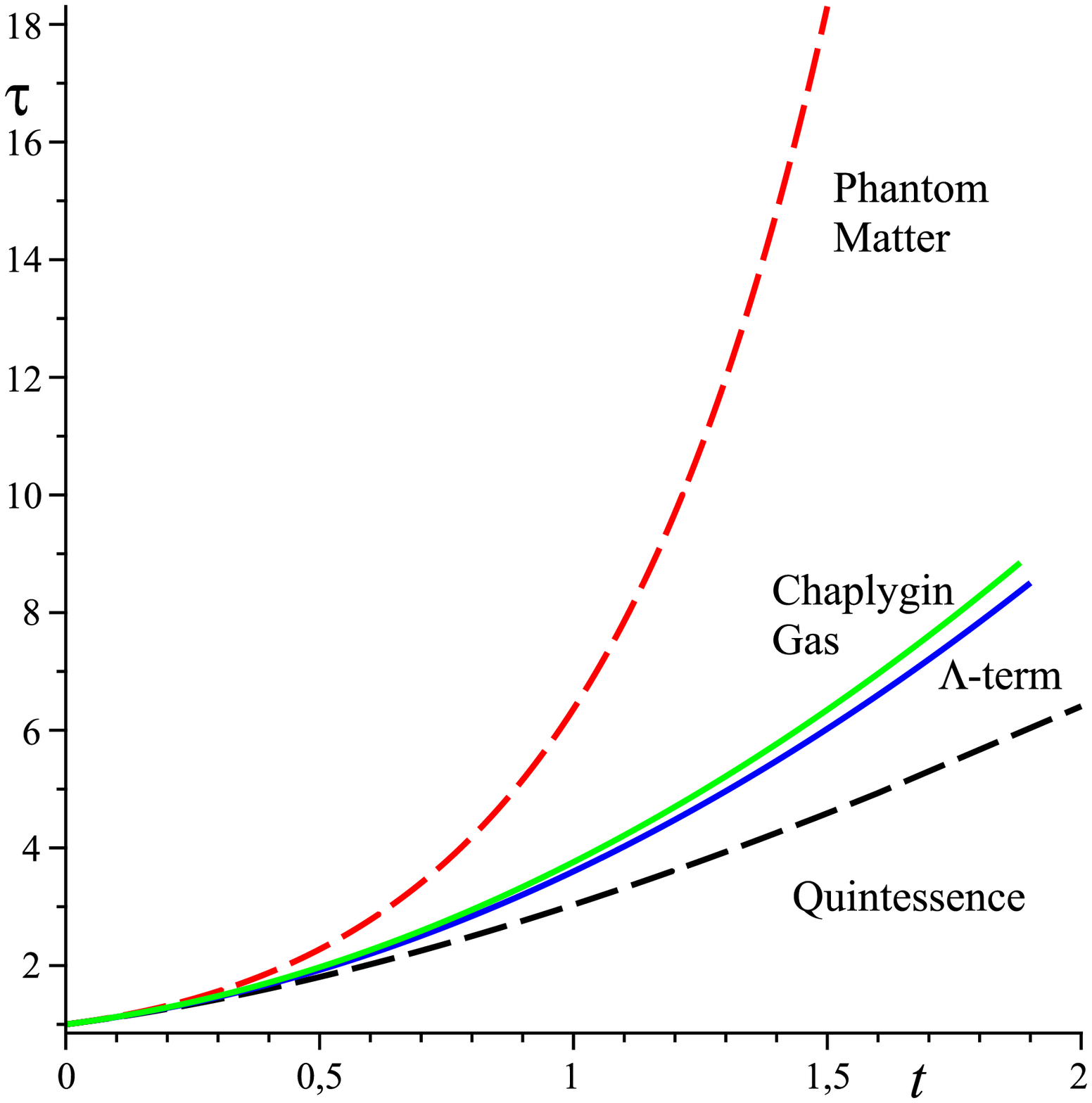}{0.45}{Evolution of the Universe
filled with dark energy.}{0.45}

In the Figs. \ref{spinpf_pf1} and \ref{spinpf_de1} we have plotted
the evolution of the Universe defined by the nonlinear spinor field
corresponding to perfect fluid and dark energy.

Let us consider the case when the spinor field is given by the
Lagrangian \eqref{lspin2}. The equation for $\tau$ now reads
\begin{equation}
\ddot \tau = (3/2) \kappa \Biggl[ \bigl(A\tau^{1+\gamma} + \lambda
C_0^{1+\gamma}\bigr)^{1/(1+\gamma)} + A
\tau^{1+\gamma}/\bigl(A\tau^{1+\gamma} + \lambda
C_0^{1+\gamma}\bigr)^{\gamma/(1+\gamma)}\Biggr],
\end{equation}
with the solution
\begin{equation}
\frac{d \tau}{\sqrt{C_1 + 3 \kappa \tau \bigl(A\tau^{1+\gamma} +
\lambda C_0^{1+\gamma}\bigr)^{1/(1+\gamma)}}} = t + t_0, \quad C_1 =
{\rm const}. \quad t_0 = {\rm const}.
\end{equation}
Inserting $\gamma = 1$ we come to the result obtained in
\cite{chjp}.

Finally we consider the case with modified quintessence. In this
case for $\tau$ we find
\begin{equation}
\ddot \tau = (3/2) \kappa \Bigl[\eta C_0^{1-W} (1+W) \tau^{W} - 2W
\ve_{\rm cr}\tau/(1-W)\Bigr],
\end{equation}
with the solution in quadrature
\begin{equation}
\frac{d\tau}{\sqrt{3 \kappa \bigl[\eta C_0^{1-W} \tau^{1+W} -
W\ve_{\rm cr}\tau^2/(1-W)\bigr]  + C_1}} = t + t_0.
\end{equation}
Here $C_1$ and $t_0$ are the integration constants.

\myfigures{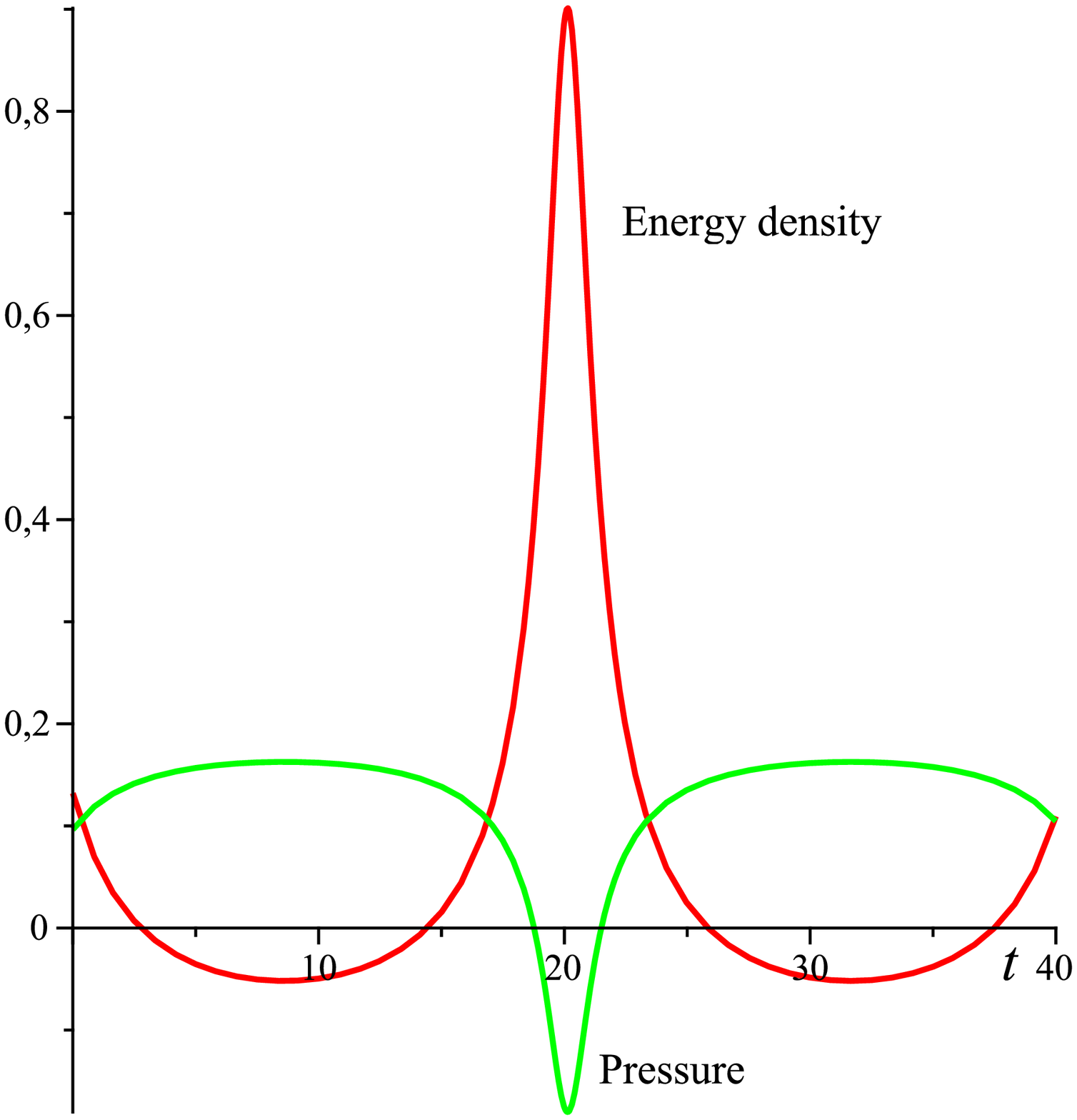}{0.45}{Dynamics of energy density and
pressure for a modified quintessence.}
{0.45}{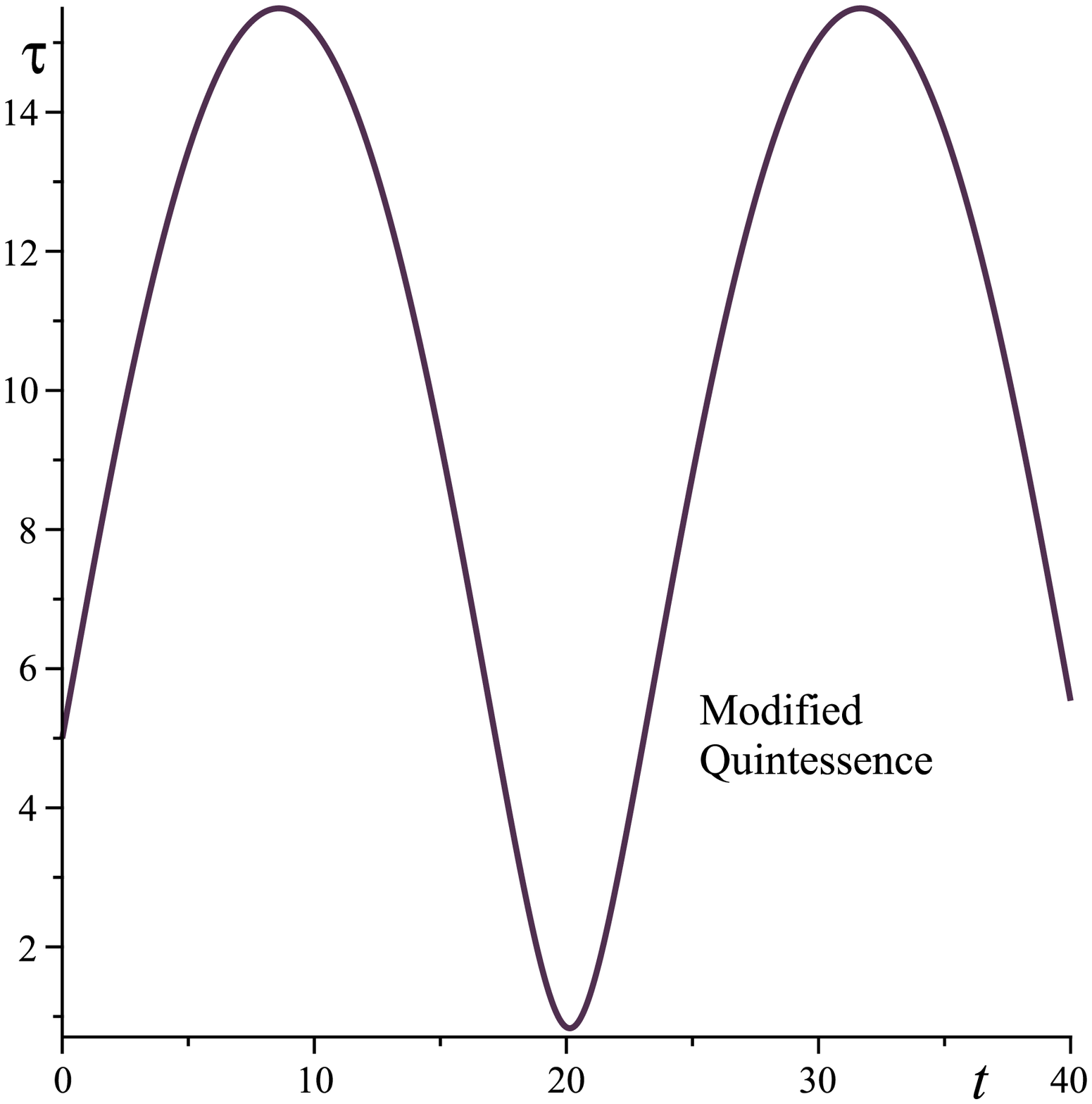}{0.45}{Evolution of the Universe filled with a
modified quintessence.}{0.45}

In the Fig. \ref{spinpf_mqep1} we have illustrated the dynamics of
energy density and pressure of a modified quintessence. In the Fig.
\ref{spinpf_mq1} the evolution of the Universe defined by the
nonlinear spinor field corresponding to a modified quintessence has
been presented. As one sees, in the case considered, acceleration
alternates with declaration. In this case the Universe can be either
singular (that ends in Big Crunch) or regular.

\section{Conclusion}

Within the framework of cosmological gravitational field equivalence
between the perfect fluid (and dark energy) and nonlinear spinor
field has been established. It is shown that different types of dark
energy can be simulated by means of a nonlinear spinor field. Using
the new description of perfect fluid or dark energy one can study
the evolution of the Universe.

\end{document}